\documentclass[aps,pre,twocolumn,floatfix,superscriptaddress,showpacs,showkeys,nofootinbib]{revtex4-1}
\usepackage{epsf,amsmath,amssymb,verbatim,color,multirow,pifont}
\usepackage{graphicx}
\usepackage{braket}
\usepackage{hyperref}
\usepackage{cleveref}

\usepackage[utf8]{inputenc}

\DeclareMathOperator{\for}{for}
\DeclareMathOperator{\otherwise}{otherwise}
\newcommand{\Matrix}[1]{\textit{\textbf{#1}}}

\newcommand{\Order}[2]{\mathcal{O}(#1^{#2})}

\crefname{equation}{Eq.}{Eqs.}
\crefname{section}{Sec.}{Secs.}
\crefname{figure}{Fig.}{Figs.}
\crefname{table}{TABLE}{Figs.}

\begin{document}
\setcounter{page}{1}
\title[]{Fast algorithm for relaxation processes in big-data systems}
\author{S.~\surname{Hwang}}
\affiliation{Department of Physics and Astronomy, Seoul National University, Seoul 151-747, Korea}
\author{D.-S.~\surname{Lee}}
\email{deoksun.lee@inha.ac.kr}
\affiliation{Department of Physics and Department of Natural Medical Sciences, Inha University, Incheon 402-751, Korea}
\author{B.~\surname{Kahng}}
\email{bkahng@snu.ac.kr}
\affiliation{Department of Physics and Astronomy, Seoul National University, Seoul 151-747, Korea}
\date[]{Received \today}

\begin{abstract}
Relaxation processes driven by a Laplacian matrix can be found 
	in many real-world big-data systems, for example,
	in search engines on the World-Wide-Web and the dynamic load balancing 
	protocols in mesh networks. 
To numerically implement such processes, a fast-running algorithm for 
	the calculation of the pseudo inverse of the Laplacian matrix is essential. 
Here we propose an algorithm which computes fast and 
	efficiently the pseudo inverse of Markov chain generator matrices
	satisfying the detailed-balance condition, a general class of matrices 
	including the Laplacian. 
The algorithm utilizes the renormalization of the Gaussian integral. 
In addition to its applicability to a wide range of problems,
	the algorithm outperforms other algorithms 
	in its ability to compute within a manageable computing time arbitrary 
	elements of the pseudo inverse of a matrix of size millions by millions. 
Therefore our algorithm can be used very widely in analyzing the relaxation 
processes occurring on large-scale networked systems.
\end{abstract}


\maketitle

\section{Introduction}
\label{sec:intro}

Fast analyses of big datasets~\cite{SNAP} are increasingly requested in diverse interdisciplinary area in this information era. Given the limitations of available computing resources in space and time, designing and implementing scalable and efficient algorithms are essential for practical applications.
One of the tasks most often encountered in such problems is the analysis of huge sparse matrices, for example, the Laplacian matrix $\Matrix{L}$ of a large-scale complex network.
The matrix $\Matrix{L}$ plays important roles in a wide range of problems such as diffusion processes, random walks~\cite{Noh2004,Meroz2011},  search engines on web pages \cite{Langville2006}, synchronization phenomena~\cite{Barahona2002}, epidemics~\cite{Wang2003}, and load balancing in parallel computing~\cite{Korniss2000}.
For instance, the spectrum of a Laplacian matrix determines the number of 
minimum spanning tree, minimal cuts~\cite{Biggs1994,Chung1996} and Kirchhoff 
index~\cite{Bonchev1994}.

The elements of the Laplacian matrix $\Matrix{L}$ of a given network are 
represented as ${L}_{ij} = \delta_{ij}-{A}_{ij}/k_j$  with the degree of node 
$j$ given by $k_j =\sum_\ell {A}_{j\ell}$. The Laplacian matrix has a couple 
of  remarkable features. It has positive eigenvalues and one non-degenerate 
zero eigenvalue. The zero eigenvalue appears since  $\sum_i \Matrix{L}_{ij}=0$, 
related to e.g., the probability conservation in the context of random walks 
and diffusion. 
Also, the Laplacian matrix can be symmetrized as
	$\bar{\Matrix{L}} = \Matrix{S}\Matrix{L}\Matrix{S}^{-1}$, whose element is 
	given as $\bar{L}_{ij} = \delta_{ij} - {A}_{ij}/\sqrt{k_i k_j}$ for $S_{ij} 
	= k_i^{-1/2} \delta_{ij}$. 
This symmetrization can be performed not only for the Laplacian matrix 
	but also for all the generators $\Matrix{V}$ of Markov chains 
	satisfying the detailed-balance condition~\cite{Kampen1992}, 
	the definition of which will be explained in detail later.
In this paper, we propose an algorithm for 
	computing the generalized inverse, so-called Moore-Penrose pseudo inverse 
	of those generators, which is relevant to the first passage property and 
	the correlation function of the Markov chains and 
therefore has been extensively studied in the physics context~\cite{Meyer1975,Rising1991,Ben-Israel2010,Kemeny1983,Hunter2014}. 

If one uses the standard eigendecomposition method based 
	on the QR algorithm \cite{Press2007}, 
	it needs $\Order{N}{2}$ memory space and takes $\Order{N}{3}$ computing 
	time to obtain the inverse of a  $N \times N$ matrix.
Therefore this algorithm cannot be actually applied 
	to obtain the inverse of large-size matrices and faster algorithms have 
	been developed to solve specific problems handling large sparse matrices. 
For instance, the iterative methods such as the well-known Jacobi method 
	or the Krylov subspace method~\cite{Saad2003} are very efficient for the 
	linear problem $\Matrix{M}\ket{x}=\ket{b}$.
In the Euclidean lattice, the Fourier acceleration method has been introduced 
	to overcome slow convergence of the Jacobi method for random resistor 
	networks embedded in the Euclidean 
	space~\cite{Batrouni1986,Batrouni1987,Batrouni1988}. 
Also the graph theoretic methods such as the fast inverse using nested 
dissection(FIND) are known to be efficient for computing the inverse of large 
sparse positive-definite 
matrices~\cite{George1973,George1981,Li2008,Lin2011,Eastwood2013},
	most of which are useful in two-dimensional lattice.
The pseudo inverse of singular matrices have been 
	investigated~\cite{Ho2005,Vishnoi2013,Ranjan2013} 
	and can be obtained efficiently for each specific domain of strength 
	such as for bipartite graphs\cite{Ho2005}, 
	the linear problem of the Laplacian matrix~\cite{Vishnoi2013}, 
	or the Laplacian-specific method~\cite{Ranjan2013}.

Our algorithm can be used for a wide range of problems effectively;
	it enables to obtain a set of $\mathcal{O}(N)$ arbitrary 
	elements of the pseudo inverse of a class of $N\times N$ matrices  
	within the computing time much shorter than $\mathcal{O}(N^3)$ in most 
	cases.
Note that the solution to a single linear problem cannot provide a set of 
arbitrary elements of the pseudo inverse in a single run.
The class of the singular matrices we consider here are the generators 
of  the Markov chains satisfying the detailed-balance condition.
The algorithm exploits the fact that the Gaussian integral with a coupling 
matrix $\Matrix{H}$ under external fields turns into a Gaussian function  of 
the external-field variables with  the coupling matrix given by 
$\Matrix{H}^{-1}$. 
The coupling matrix $\Matrix{H}$ is constructed from a given generator matrix 
$\Matrix{V}$.
Its Gaussian integral is evaluated by decimating the 
variables and renormalizing the coupling matrix  
	with an appropriate treatment of the  zero eigenvalue mode of $V$.  
  
To verify the usefulness and performance of the proposed algorithm in physics 
problems,  we apply  the algorithm to compute the global mean first passage 
time (GMFPT) of random walk on various networks, which requires the computation 
of all the diagonal elements of  the pseudo inverse of the generator - the 
Laplacian matrix.
We compare the computational cost of our algorithm with 
that of the QR algorithm for small system sizes and that of  the random-walk simulation.  
The dependence of network topology on the computing time of our algorithm 
is discussed.

This paper is organized as follows. In Sec.~\ref{sec:gaussian}, 
	we introduce the basic formulae of the  Gaussian integral, 
	which play the central roles in designing our algorithm.
Before presenting the main algorithm, the one 
computing the inverse of a positive-definite matrix is outlined in 
Sec.~\ref{sec:outline}. 
In Sec.~\ref{sec:target}, we specify a target problem of our algorithm,
	of which the pseudo inverse can be obtained exactly by our algorithm.
The applications of the pseudo inverse 
	in the physics context are also presented. 
In Sec.~\ref{sec:algorithm}, we describe each procedure of the algorithm in 
detail.
In \cref{sec:performance},  the running time of our algorithm to compute the GMFPT on various model networks is presented and compared  with that of other methods. 
The  algorithm is applied  to large real-world networks, demonstrating its practical use in the same section. The summary and discussion are given in Sec.~\ref{sec:summary}.

\section{Basic Formulation}
\label{sec:gaussian}

Here we present the formulae of which will be used in our algorithm.
For an $N \times N$ non-singular real symmetric matrix $\Matrix{H}$ and an 
arbitrary 
column vector $\ket{J}$ of size $N$, we consider the Gaussian integral given 
by  
\begin{align}
Z \equiv& \int_{-\infty}^{\infty} \prod_{j=1}^N {\mathrm d}\phi_j \exp
\left[
	\frac{i}{2} \bra{\phi} \Matrix{H} \ket{\phi}  + i \braket{J | \phi}
\right] \nonumber \\
=& \sqrt{\frac{(2 \pi i)^{N}}{\det \Matrix{H}}} e^{-\frac{i}{2}\bra{J}{\Matrix{H}}^{-1}\ket{J}},
\label{eq:gaussian}
\end{align}
where $|\phi\rangle = (\phi_1, \phi_2, \ldots, \phi_N)^\dagger$  and  the factor $i$'s are introduced for the convergence of the integral. Once the Gaussian integral $Z$ is evaluated,  the inverse matrix ${\Matrix{H}}^{-1}$ can be obtained by 
\begin{align}
{H}^{-1}_{j\ell} & = -i \left. \frac{\partial^2}{\partial J_j \partial J_\ell} \log Z \right|_{\ket{J} = \ket{0}},
\label{eq:inverse}
\end{align}
where $\ket{0}$ is a null vector.
If we introduce a $2N$-dimensional vector $\ket{\psi}$ by gluing  $\ket{J}$ and $\ket{\phi}$ as
\begin{align}
\psi_{j} =
\left \{
\begin{array}{cc}
J_{j} & \for ~~ 1 \le j \le N,\\ 
\phi_{j-N} & \for ~~ N + 1 \le j \le 2N,\\ 
\end{array} 
\right. 
\label{eq:psi}
\end{align}
and  a $2N \times 2N$ real symmetric matrix $\tilde{\Matrix{H}}$ 
\begin{align}
\tilde{ H}_{j\ell}=\left \{
	\begin{array}{cc}
		\delta_{j \ell} & \for ~~ 1 \le  j, \ell \le N,\\ 
		{H}_{j-N, \ell-N} & \for ~~ N+1 \le  j, \ell \le 2N,\\ 
		0 & \otherwise,
	\end{array} 
\right. 
\label{eq:extension}
\end{align}
we can represent Eq.~(\ref{eq:gaussian}) in a simple form as
\begin{align}
Z  = \int_{-\infty}^{\infty} \prod_{\ell=N+1}^{2N} {\mathrm d}\psi_\ell \exp\left[ \frac{i}{2} \bra{\psi} \tilde{\Matrix{H}} \ket{\psi}\right].
\label{eq:gaussian2}
\end{align}

The evaluation of the Gaussian integral in Eq.~(\ref{eq:gaussian2}) can be done by integrating out $\psi$ variables one by one and  renormalizing the elements of $\tilde{\Matrix{H}}$ accordingly. The matrix $\tilde{\Matrix{H}}$ thus reduces its dimension by one at every stage. Some of the zero elements in $\tilde{\Matrix{H}}$ can be nonzero after such renormalization,  which should be taken care of as detailed in the next section. For an extended coupling matrix $\tilde{\Matrix{H}}$, we consider a graph  $G$ with the adjacency matrix $\Matrix{A}$ with its elements given by 
\begin{align}
{A}_{j\ell} = \left \{
\begin{array}{ll}
1 & \ {\rm if} \ \tilde{H}_{j\ell}\neq 0, \\
0 & \ {\rm otherwise} 
\end{array} 
\right. 
\label{eq:adj}
\end{align}
$G$ and $\Matrix{A}$ then evolve as $\tilde{\Matrix{H}}$ is renormalized successively.

\section{Outline of the algorithm for the inverse of a positive-definite matrix}
\label{sec:outline}

In this section, we outline the algorithm for computing the inverse of a positive definite matrix $\Matrix{H}$ 
by evaluating the Gaussian integral in Eq.~(\ref{eq:gaussian2}), 
 which will be generalized to singular matrices in Sec.~\ref{sec:algorithm}. 
 For an $N\times N$ positive definite matrix $\Matrix{H}$, following Eq.~(\ref{eq:extension}), we construct the extended matrix $\tilde{\Matrix{H}}$ of $2N\times 2N$. The corresponding graph $G$ of $2N$ vertices  has the adjacency matrix $\Matrix{A}$ as in Eq.~(\ref{eq:adj}).

\begin{figure}
\includegraphics[width=8.5cm]{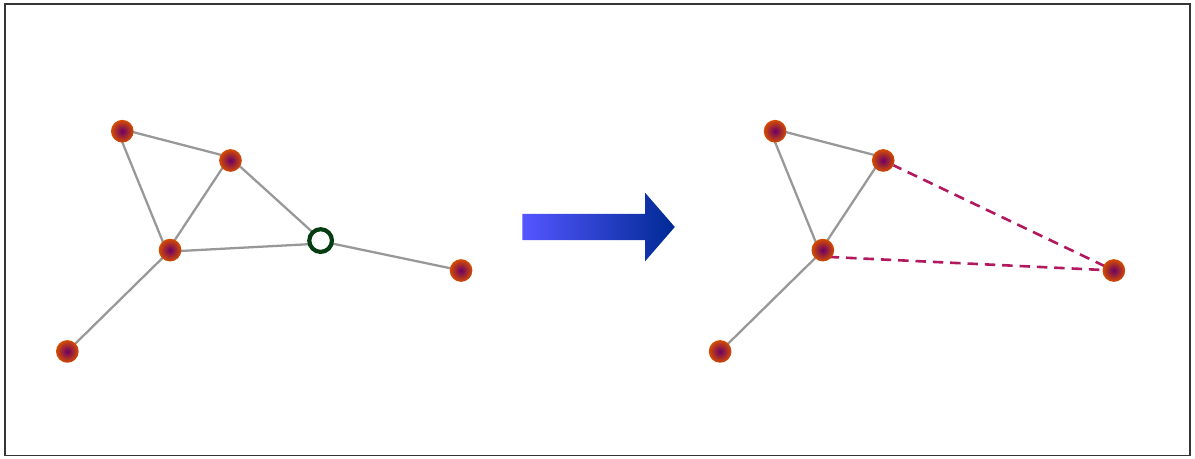}
\caption{(Color Online) Example of eliminating a node in a graph.  
When a node (open circle) is eliminated, new links (dashed lines) are added to 
	the pairs of its neighbor nodes if disconnected. 
Consequently, the neighbor nodes form a clique, 
	a completely-connected subgraph.   
\label{diagram}
}
\end{figure}

Suppose that we integrate out $\psi_{2N}$ in \cref{eq:gaussian2} to transform $\tilde{\Matrix{H}}$ 
into  $\tilde{\Matrix{H}}^{(1)}$ of size $(2N-1)\times (2N-1)$. 
Since $\Matrix{H}$ is positive definite,  $\tilde{H}_{2N\, 2N}$ is positive.
Collecting the terms involving $\psi_{2N}$, we find that
\begin{align}
\int_{-\infty}^{\infty} {\mathrm d} \psi_{2N} & \exp \left( \frac{i}{2} \tilde{{H}}_{2N\, 2N} \psi_{2N}^2 + i B_{2N}\psi_{2N} \right) \nonumber\\
&= \sqrt{\frac{2 \pi i}{ \tilde{{H}}_{2N\, 2N} }} \exp\left(-\frac{i}{2} \frac{B_{2N}^2} { \tilde{{H}}_{2N\, 2N} } \right),
\label{eq:deleted}
\end{align}
where $B_{2N}\equiv \sum_j  A_{j\, 2N}\tilde{{H}}_{j\, 2N} \psi_j $. Noting that 
 $B_{2N}^2 = \sum_{j,\ell}{A}_{j\, 2N} {A}_{\ell\,  2N} \tilde{ H}_{j\, 2N} \tilde{ H}_{\ell\, 2N} \psi_j \psi_\ell$, one can identify the renormalized Hamiltonian $\tilde{\Matrix{H}}^{(1)}$ in the Gaussian integral as 
\begin{align}
Z &= \sqrt{\frac{2 \pi i}{ \tilde{{H}}_{11} }} Z^{(1)}, \nonumber\\
Z^{(1)} &= \int_{-\infty}^{\infty} \prod_{j=N+1}^{2N-1} {\mathrm d}\psi_j \exp
\left[
	\frac{i}{2} \bra{\psi} \tilde{\Matrix{H}}^{(1)} \ket{\psi}
\right],
\label{eq:renormalize}
\end{align}
where $\tilde{H}^{(1)}_{j\ell} = \tilde{H}_{j\ell}$ for all $1\leq j,\ell \leq (2N-1)$ unless both $j$ and $\ell$ are the neighbor nodes of the decimated node $2N$ in $G$, the graph representation of $\tilde{\Matrix{H}}$.
If ${A}_{j\, 2N} {A}_{\ell\, 2N}>0$,  the corresponding matrix element $\tilde{H}_{j\ell}$ is changed to $\tilde{H}^{(1)}_{j\ell} =\tilde{{H}}_{j\ell} - \tilde{{H}}_{j\, 2N} \tilde{{H}}_{\ell\, 2N}/{\tilde{{H}}_{2N\, 2N}}$.
In the graph representation, the corresponding graph $G$ is transformed to $G^{(1)}$ by eliminating the node $2N$ and adding links to every pair of the nodes  that were adjacent to the node $2N$ but disconnected in $G$ (see Fig.~\ref{diagram}). Accordingly, the adjacency matrix evolves from ${\bf A}$ to ${\bf A}^{(1)}$. 

We repeat this procedure, decimation followed by renormalization,  $N$ times to  integrate out all $\psi_{j}=\phi_{j-N}$ variables for $N+1\leq j\leq 2N$. Consequently the extended matrix $\Matrix{H}$ evolves as 
\begin{equation}
\tilde{\Matrix{H}}^{(0)} = \tilde{\Matrix{H}}\to \tilde{\Matrix{H}}^{(1)}\to \cdots \to \tilde{\Matrix{H}}^{(N-1)}\to \tilde{\Matrix{H}}^{(N)}=\Matrix{H}^{-1}
\end{equation}
 while reducing its dimension from $2N$ to $N$. The $N\times N$ matrix $\tilde{\Matrix{H}}^{(N)}$ obtained at the last stage represents the coupling between $J$'s and is equal to $\Matrix{H}^{-1}$ in Eq.~(\ref{eq:gaussian}). The adjacency matrix $\Matrix{A}$ and the graph $G$ also evolve as $\Matrix{A} = \Matrix{A}^{(0)} \to \Matrix{A}^{(1)} \to \cdots \to \Matrix{A}^{(N)}$ and $G=G^{(0)}\to G^{(1)}\to\cdots \to G^{(N)}$, respectively.

The order of decimating nodes affects significantly  the running time of the algorithm, and  will be discussed in detail later. Once it is determined, one can rearrange the node indices of $\tilde{\Matrix{H}}$ such that nodes are eliminated from  $j=2N$ to $j=N+1$. That is,  if we introduce $v_n$, the index of the node that is eliminated in $G^{(n)}$ to obtain $G^{(n+1)}$,  it is given by $v_n = 2N-n$ for $n=0, 1, \ldots, N-1$. The matrix $\tilde{ \Matrix{H}}^{(n+1)}$ is obtained by removing the last row and column of $\tilde{\Matrix{H}}^{(n)}$, corresponding to the node $v_{n}=2N-n$, and updating the elements $\tilde{H}_{j\ell}$ for both $j$ and $\ell$ adjacent to $v_n$ as  
\begin{equation}
\tilde{{H}}^{(n+1)}_{j\ell} = \tilde{{H}}^{(n)}_{j\ell} - {A}^{(n)}_{j v_n} {A}^{(n)}_{\ell v_n} \frac{ \tilde{{H}}^{(n)}_{j v_n} \tilde{{H}}^{(n)}_{\ell v_n}}{\tilde{{H}}^{(n)}_{v_n v_n}}
\label{eq:coefUpdate}
\end{equation}
for $1\leq j,\ell\leq v_{n+1}=v_n-1$.
We remark that $\tilde{{H}}^{(n)}_{v_n v_n}\ne 0$ and therefore one can apply 
Eq.~(\ref{eq:coefUpdate}) for $n=0,1,2,\ldots, N-1$. This can be understood as 
follows:
If $\tilde{{H}}^{(n)}_{v_n v_n}=0$ for $0\leq n<N$, the $(n+1)\times (n+1)$ 
block matrix $\Matrix{B}$ representing the coupling among $v_0, v_1, \cdots, 
v_n$ should have its determinant equal to zero, since $\det \Matrix{B} \propto  
\prod_{\ell=0}^{n} \tilde{{H}}^{(\ell)}_{v_\ell v_\ell}$ as shown in 
\cref{eq:gaussian}, and the latter is zero under the assumption that 
$\tilde{{H}}^{(n)}_{v_n v_n}=0$. This contradicts the condition that 
$\Matrix{H}$ is positive definite since all sub-matrices of a positive definite 
matrix are also positive definite. 

The adjacency matrix $\Matrix{A}^{(n+1)}$ is obtained by removing the last row and column of $\Matrix{A}^{(n)}$ and updating the element as  
\begin{equation}
\Matrix{A}_{j\ell}^{(n+1)} = \Matrix{A}_{j\ell}^{(n)} + \Matrix{A}^{(n)}_{j v_n} \Matrix{A}^{(n)}_{\ell v_n} (1-\Matrix{A}^{(n)}_{j\ell}).
\label{eq:adjacencyevolution}
\end{equation}
Note that connecting each pair of the neighbor nodes of the decimated node may increase the mean degree $\langle k\rangle=2L/N$, the ratio of the number of links ($L$) to the number of nodes ($N$), of the evolving graph.  

$\Matrix{H}^{(N)}$  is uniquely determined regardless of the order of decimating nodes.  However, the ordering $v_n$ is important for reducing the computational cost. For instance, as shown in Fig.~\ref{diagram} and Eq.~(\ref{eq:adjacencyevolution}), if a node with degree $k$ is removed,  its $k$ links are removed but its neighbors get interconnected, resulting in the maximum possible increase of links by  $k(k-1)/2 - k$: If a hub node is eliminated, one should update lots of elements $\tilde{H}_{j\ell}$'s  according to Eq.~(\ref{eq:coefUpdate}) in the following stages of evolution, which increases the computing time.   

The appearance of new links as in Fig.~\ref{diagram} and Eq.~(\ref{eq:adjacencyevolution}) are called {\it fill-in} in the context of graph theory and there have been much efforts to find the optimal ordering that suppresses those fill-ins. Eliminating nodes in a graph, so called {\it graph elimination game}, is encountered in the Cholesky factorization,  which is generally used to solve linear problem $\Matrix{M}\ket{x} = \ket{b}$ for positive definite matrix $\Matrix{M}$. 
While the ideal ordering which minimizes the fill-ins is hard to find, 
heuristic methods have been proposed, such as the minimum-degree ordering, the reverse Cuthill-McKee ordering, and the nested-dissection ordering~\cite{George1981,George2011}. 

In decimating $\phi_j (=\psi_{j-N})$ variables in Eq.~(\ref{eq:gaussian2}),  every pair of nodes  that are adjacent to the decimated node should update their corresponding matrix element.  If we classify  
the pairs of nodes $(j \ell)$ into three groups according to the types of their associated variables as  ($\phi_j\phi_\ell$), ($J_j J_\ell$) and ($\phi_j J_\ell$), the computing time is expected to increase  if many fill-ins appear  for pairs of type $(\phi,J)$ or $(\phi,\phi)$.
Therefore, we here choose the minimum-degree ordering which minimizes the fill-ins for $(\phi_j, \phi_\ell)$. To find the order of decimating nodes and rearrange the node indices so that nodes are  eliminated from  the one with $j=2N$ to $N+1$ in $\tilde{\Matrix{H}}$ after the rearrangement,  we perform the node elimination in $G$ representing $\Matrix{H}$ as follows:
\begin{enumerate}
\item Construct graph $G^{(0)} = G$ representing $\Matrix{H}$.
\item $n \gets 0$.
\item Choose one of the nodes having the minimum degree in $G^{(n)}$ and record its index in $w(n)$.
\item Assign a link to every disconnected pair of neighboring nodes of the node $w(n)$ and  eliminate the node $w(n)$ and its links, which yields $G^{(n+1)}$. 
\item If $n<N$, $n \gets n+1$ and go to the step 3. Otherwise,  for each node 
of index $i = 1,2, \ldots, N$ of $\Matrix{H}$, assign a new index $N-w(i)$. 
\end{enumerate}

\section{Target problem}
\label{sec:target}

Our idea is that one can use the method in Sec.~\ref{sec:outline} to obtain a set of the arbitrary elements of the pseudo inverse of an $N\times N$ matrix $\Matrix{V}$ satisfying the following conditions:
\begin{enumerate}
\item  \Matrix{V} is a semi-positive definite symmetric matrix with the zero eigenvalue $\lambda_1=0$ of multiplicity 1, and 
\item has the  eigenvector $\ket{\Matrix{e}^{(1)}}$ corresponding to the zero eigenvalue, which  does not have any zero component, i.e., ${ e}_i^{(1)} \ne 0$ for $1\leq i\leq N$. 
\end{enumerate}
We call  such a matrix a semi-positive definite symmetric (SPDS) matrix 
	for simplicity.
For  a SPDS matrix $\Matrix{V}$, one can define its pseudo-inverse matrix 
$\Matrix{V}^+$ by dropping the zero-eigenvalue mode as
\begin{equation}
\Matrix{V}^+ = \sum_{n=2}^N {|\Matrix{e}^{(n)}\rangle \langle \Matrix{e}^{(n)}|\over \lambda_n},
\end{equation} 
where $\lambda_n$'s are the eigenvalues of $\Matrix{V}$ with $\lambda_1=0$ and $|\Matrix{e}^{(n)}\rangle$'s are the corresponding eigenvectors. 

The generator  $\Matrix{V}$ of a Markov chain satisfying the detailed-balance condition is an example \cite{Kampen1992}.  
$\Matrix{V}$ has the zero eigenvalue of multiplicity 1:  Its left eigenvector is $\langle \Matrix{e}^{(1)}|=(1,1, \cdots, 1)$, representing the conservation of the probability,   and 
the component $e_j^{(1)}$ of the right eigenvector $|\Matrix{e}^{(1)}\rangle=(e_1^{(1)}, e_2^{(0)},\ldots)^\dagger$ represents the stationary-state probability of the state $j$. The detailed-balance condition requires that the transition from a state $i$ to another $j$ happens with equal probability to that of the transition from $j$ to $i$. If $\Matrix{V}$ satisfies the detailed-balance condition, all the components $e_j^{(1)}$'s should be nonzero.  If $\Matrix{V}$ is not symmetric,  a symmetric matrix $\bar{\Matrix{V}}$ can be obtained by the similarity transformation 
\begin{align}
\bar{\Matrix{V}}=\Matrix{S}\Matrix{V}\Matrix{S}^{-1}
\end{align}
 with ${S}_{j\ell} = \delta_{j\ell}/\sqrt{e^{(1)}_j} $. The matrix $\bar{\Matrix{V}}$ is then a SPDS matrix.
To obtain the stationary-state probability, there have been lots of efficient algorithms suggested so far, e.g., see \cite{Benzi2002}.

As a concrete example of SPDS matrices, the Laplacian $\Matrix{L}$ with  $L_{j\ell}=\delta_{j\ell} - A_{j\ell}/k_\ell$ generates the time evolution of the occupation probability $P_j(t)$ of a random walker as
$P_j(t+1)=P_j(t) - \sum_\ell {L}_{j\ell} P_\ell(t)$. 
One can symmetrize $\Matrix{L}$ by the transformation $\bar{\Matrix{L}}=\Matrix{S}\Matrix{L}\Matrix{S}^{-1}$ with $S_{j\ell}=\delta_{j\ell}k_j^{-1/2}$ to obtain $\bar{L} _{j\ell}= \delta_{j\ell}-{A}_{j\ell}/\sqrt{k_j k_\ell}$. The pseudo inverse  $\Matrix{L}^+$ or $\bar{\Matrix{L}}^+$ contains important information of random walk dynamics.
For instance, the mean-first passage time (MFPT) $T_{is}$ from a node $s$ to $i$ is represented as~\cite{Chung2000,Qiu2007} 
\begin{equation}
T_{is} = \left\{
\begin{array}{ll}
{2L\over k_i}\left({L}^+_{ii} - {L}^+_{is}\right) ={2L\over k_i}\left(\bar{{L}}^+_{ii} -\sqrt{k_i\over k_s}\bar{ {L}}^+_{is}\right) & \ {\rm for} \  i\ne s,\\
{2L \over k_i} &\ {\rm for} \ i = s.
\end{array}
\right.
\label{eq:Tis}
\end{equation}
The GMFPT $T_i$ of node $i$ denotes the MFPT to the target node $i$ averaged over all possible starting nodes in the stationary state~\cite{Hwang2012a} and is represented by the diagonal element of the pseudo inverse of the Laplacian as
\begin{equation}
T_i = \sum_s {k_s\over 2L} T_{is} = {2L\over k_i} {L}^+_{ii}+1 = {2L\over k_i} \bar{{L}}^+_{ii}+1.
\label{eq:GMFPT}
\end{equation}

Another Laplacian $\hat{\Matrix{L}}$ with its element given by $\hat{{L}}_{j\ell} = k_j \delta_{j\ell} - {A}_{j\ell}$
is the time-evolution operator of  the Edwards-Wilkinson model describing the fluctuating interfaces under tension and noise as  $\dot{h_j} = -\sum_\ell \hat{{L}}_{j\ell} h_\ell +\xi_j(t)$ with $h_j$ the height at site $j$ and $\xi_j(t)$ the noise~\cite{Barabasi1995}.
The   height-height correlation is represented in terms of the pseudo inverse of $\hat{\Matrix{L}}$ as 
\begin{equation}
\langle (h_j-\bar{h}) (h_\ell-\bar{h}) \rangle=\hat{{L}}^{+}_{j\ell}
\label{eq:hij}
\end{equation} 
with the mean height $\bar{h}=N^{-1}\sum_j h_j$~\cite{Korniss2000,Kozma2004,Guclu2007}. The roughness is defined as $w = \sqrt{N^{-1}\sum_j\langle (h_j -\bar{h})\rangle^2}$ and is evaluated by
\begin{equation}
w =\sqrt{{1\over N} \sum_j \hat{{L}}^+_{jj}}.
\label{eq:roughness}
\end{equation}

As shown above, the MFPT and the GMFPT of random walk and the height-height correlation and the roughness of fluctuating interfaces are commonly represented in terms of 
$\mathcal{O}(N)$ number of elements of the  pseudo inverse of an $N\times N$ 
SPDS matrix.
The algorithm in the next section is appropriate for computing a 
	set of such {\it arbitrary} elements of the pseudo inverse of a SPDS 
	matrix. 
Other algorithms are optimal for $N$ small~\cite{Qiu2007}, for the linear problems~\cite{Saad2003,Batrouni1986,Batrouni1987,Batrouni1988,Vishnoi2013}, for the positive-definite matrices~\cite{George1973,George1981,Li2008,Lin2011,Eastwood2013}, or for  limited cases~\cite{Ho2005,Ranjan2013}. 
The linear problem $\Matrix{M}\ket{x}=\ket{b}$  is encountered in numerous 
applications and can give for instance the $k$-th column of $\Matrix{M}^+$ by 
setting $b_j=\delta_{jk}$. However, the solution to such a single linear 
problem cannot give the sum of the diagonal elements $\sum_{j} M^+_{jj}$ as 
required in the GMFPT or the roughness. In contrast, our algorithm obtains a 
set of $\mathcal{O}(N)$ {\it arbitrary} elements of the pseudo inverse of 
a large SPDS matrix at a time. In general,  the inverse of a sparse matrix is 
not guaranteed to be sparse. Therefore given the limitation of space and time 
of computation,  it is not always available to obtain {\it all} the elements of 
the inverse matrix of a large sparse matrix.

\section{Algorithm for the arbitrary elements of the pseudo inverse of a SPDS 
matrix}
\label{sec:algorithm}
Here we present the algorithm for computing the arbitrary elements of the pseudo inverse of a SPDS matrix $\Matrix{V}$. Since $\Matrix{V}$ is not invertible, we introduce $\Matrix{H}(\mu)\equiv \mu \Matrix{I} + \Matrix{V} $ where $\mu$ is a positive real constant and $\Matrix{I}$ is the identity matrix of the same dimension as $\Matrix{V}$. Then $\Matrix{H}(\mu)$ is positive definite and therefore we can apply Eqs.~(\ref{eq:gaussian}) and (\ref{eq:inverse}) to obtain 
\begin{align}
\Matrix{H}^{-1}_{j\ell}(\mu)  & = - i \left. \frac{\partial^2}{\partial J_j \partial J_\ell} \log Z(\mu) \right|_{\vec{J} = 0} \nonumber \\
& = \sum_{n=1}^{N} \frac{{e}^{(n)}_{j} { e}^{(n)}_{\ell} }{\mu + \lambda_n} \nonumber \\
& = \frac{{e}^{(1)}_{j} { e}^{(1)}_{\ell} }{\mu} + \sum_{n=2}^{N} \frac{{e}^{(n)}_{j} {e}^{(n)}_{\ell} }{\lambda_n} + \Order{\mu}{1},
\label{eq:inverseSingular}
\end{align}
where $\lambda_n (n=1,2,\ldots,N)$ are the eigenvalues of $\Matrix{V}$ and ${\bf e}^{(n)}$'s are the corresponding eigenvectors  ${\bf e}^{(n)}=({e}_1^{(n)}, {e}_2^{(n)}, \ldots, { e}_N^{(n)})^\dagger$. 
Therefore, the pseudo inverse $\Matrix{V}^{+}$ of $\Matrix{V}$ can be obtained by using $\Matrix{H}^{-1}$ 
of Eq.~(\ref{eq:inverseSingular}) as 
\begin{equation}
{V}^{+}_{j\ell} = \left. \frac{\partial}{\partial \mu}\mu {H}^{-1}_{j\ell}(\mu) \right|_{\mu=0}.
\label{eq:pseudoInverse}
\end{equation}

Equation~(\ref{eq:pseudoInverse}) implies that one can obtain $\Matrix{V}^+$ once $\Matrix{H}^{-1}(\mu)$ 
  is known as expanded in Eq.~(\ref{eq:inverseSingular}), which becomes available by 
the few first terms,  up to $\mathcal{O}(\mu^2)$, in the expansion of the extended matrix 
$\tilde{\Matrix{H}}$
\begin{align}
\tilde{{H}}_{j\ell}(\mu) &= \left. \left(1+ \mu\frac{\mathrm{d}}{\mathrm{d\mu}} + \frac{\mu^2}{2 !} \frac{\mathrm{d}^2}{\mathrm{d\mu^2}} \right) \tilde{{H}}_{j\ell}  \right|_{\mu = 0} + \Order{\mu}{3}\nonumber\\
&= \tilde{H}_{0,j\ell}+\tilde{H}_{1,j\ell}\mu+\tilde{H}_{2,j\ell}\mu^2 +\Order{\mu}{3}.
\label{eq:Hexpand}
\end{align}
Our idea is to apply the algorithm in Sec.~\ref{sec:outline} to  trace the evolution of the three coefficient matrices  $\tilde{\Matrix{H}}_{0}, \tilde{\Matrix{H}}_{1}$, and $\tilde{\Matrix{H}}_{2}$ in Eq.~(\ref{eq:Hexpand}). Using  \cref{eq:coefUpdate} and Eq.~(\ref{eq:Hexpand}), we eliminate a node with index $v_n = 2N -n$ and renormalize the coefficients at each stage $0\leq n<N-1$ as  
\begin{widetext}
\begin{eqnarray}
\tilde{ H}_{0,j\ell}^{(n+1)} &=& \tilde{ H}^{(n)}_{0,j\ell}- {A}^{(n)}_{j v_n} {A}^{(n)}_{\ell v_n} \frac{\tilde{ H}^{(n)}_{0,j v_n} \tilde{ H}^{(n)}_{0,\ell v_n}}{\tilde{ H}^{(n)}_{0,v_n  v_n}}, \nonumber\\ 
\tilde{ H}_{1,ij}^{(n+1)} &=& \tilde{ H}_{1,ij}^{(n)} -{A}^{(n)}_{j v_n} {A}^{(n)}_{\ell v_n}  \frac{ \tilde{ H}^{(n)}_{1,j v_n} \tilde{ H}^{(n)}_{0,\ell v_n} \tilde{ H}^{(n)}_{0,v_n  v_n}+\tilde{ H}^{(n)}_{0,j v_n} \tilde{ H}^{(n)}_{1,\ell v_n}
   \tilde{ H}^{(n)}_{0,v_n  v_n}-\tilde{ H}^{(n)}_{0,j v_n} \tilde{ H}^{(n)}_{0,\ell v_n} \tilde{ H}^{(n)}_{1,v_n  v_n}}{\left[\tilde{ H}^{(n)}_{0,v_n  v_n}\right]^2}, \nonumber\\ 
\tilde{ H}_{2,ij}^{(n+1)}&=&
\tilde{ H}_{2,ij}^{(n)}- {A}^{(n)}_{j v_n} {A}^{(n)}_{\ell v_n} 
\frac{ 
\tilde{ H}^{(n)}_{2,j v_n} \tilde{ H}^{(n)}_{0,\ell v_n} \left[\tilde{ H}^{(n)}_{0,v_n  v_n}\right]^2+
\tilde{ H}^{(n)}_{1,j v_n} \tilde{ H}^{(n)}_{1,\ell v_n} \left[\tilde{ H}^{(n)}_{0,v_n  v_n}\right]^2
}
{
\left[\tilde{ H}^{(n)}_{0,v_n  v_n}\right]^3
}
\nonumber\\
&-&{A}^{(n)}_{j v_n} {A}^{(n)}_{\ell v_n} \frac{ 
\tilde{ H}^{(n)}_{0,j v_n} \tilde{ H}^{(n)}_{2,\ell v_n} \left[\tilde{ H}^{(n)}_{0,v_n  v_n}\right]^2-
\tilde{ H}^{(n)}_{1,j v_n} \tilde{ H}^{(n)}_{0,\ell v_n} \tilde{ H}^{(n)}_{1,v_n  v_n} \tilde{ H}^{(n)}_{0,v_n  v_n}-\tilde{ H}^{(n)}_{0,j v_n} \tilde{ H}^{(n)}_{1,\ell v_n} \tilde{ H}^{(n)}_{1,v_n  v_n} \tilde{ H}^{(n)}_{0,v_n  v_n}}
{
\left[\tilde{ H}^{(n)}_{0,v_n  v_n}\right]^3
}
\nonumber\\
&-&{A}^{(n)}_{j v_n} {A}^{(n)}_{\ell v_n} \frac{ 
-\tilde{ H}^{(n)}_{0,j v_n} \tilde{ H}^{(n)}_{0,\ell v_n} \tilde{ H}^{(n)}_{2,v_n  v_n} \tilde{ H}^{(n)}_{0,v_n  v_n}+\tilde{ H}^{(n)}_{0,j v_n} \tilde{ H}^{(n)}_{0,\ell v_n} \left[\tilde{ H}^{(n)}_{1,v_n  v_n}\right]^2
}
{
\left[\tilde{ H}^{(n)}_{0,v_n  v_n}\right]^3
},
\label{eq:update}
\end{eqnarray} 
\end{widetext}
where 
the indices $j$ and $\ell$ run from $1$ to $v_n-1=2N-n-1$. 

Given the relation $\Matrix{H}= \mu \Matrix{I}+\Matrix{V}$ and $\Matrix{V}$ is a SPDS matrix, one eigenvalue of $\Matrix{H}$ can be zero if $\mu=0$.  During the renormalization of  $\tilde{\Matrix{H}}_0, \tilde{\Matrix{H}}_1$, and $\tilde{\Matrix{H}}_2$ as in Eq.~(\ref{eq:update}),  $\tilde{H}_{0,v_n v_n}^{(n)}$'s are non-zero for $n=0,1,2,\ldots, N-2$.  It is only in $ \tilde{\Matrix{H}}^{(N-1)}$ that a singular element $\tilde{H}^{(N-1)}_{0,v_{N-1} v_{N-1}}=0$ appears. This  can be understood similarly to in \cref{sec:outline}. Suppose that there exists $n$ such that $\tilde{H}^{(n)}_{0,v_{n} v_{n}}=\Order{\mu}{1}$ for $0\leq n < N-1$.  Then, the determinant  of the sub-matrix $\Matrix{B}$ representing the coupling among $v_0, v_1, \cdots, v_n$ should be $\Order{\mu}{1}$, as $\det \Matrix{B}\propto \prod_{\ell=0}^n \tilde{{H}}^{(\ell)}_{v_\ell v_\ell}$ and $\tilde{\Matrix{H}}^{(n)}_{v_n v_n}=\Order{\mu}{1}$ even if $\tilde{\Matrix{H}}^{(\ell)}_{v_\ell v_\ell}=\Order{1}{}$ for all $0\leq \ell<n$. 
This means that the vector space spanned  by $v_0, v_1, \ldots, v_n$ contains the eigenvector of $\tilde{\Matrix{H}}$ associated with the zero eigenvalue for $\mu=0$, which contradicts the condition that  the eigenvector of $\Matrix{V}$ and in turn that of $\tilde{\Matrix{H}}$ associated to the zero eigenvalue  has no zero component.
Therefore, $\tilde{H}^{(n)}_{0,v_{n} v_{n}}=0$ should appear only for $n=N-1$.

At the last step of decimation, when the last node $v_{N-1}=N+1$ should be eliminated, its matrix element $\tilde{H}^{(N-1)}_{v_{N-1} v_{N-1}}$ becomes zero for $\mu=0$, that is, $\tilde{H}^{(N-1)}_{v_{N-1} v_{N-1}}(\mu)=\tilde{H}^{(N-1)}_{1,v_{N-1} v_{N-1}} \mu + \tilde{H}^{(N-1)}_{2,v_{N-1} v_{N-1}} \mu^2 +\mathcal{O}(\mu^3)$. Using this expansion in Eq.~(\ref{eq:coefUpdate}), 
one can find that $\tilde{H}^{(N)}_{j\ell}$ is expanded as 
\begin{equation}
{ H}^{-1}_{j\ell}(\mu)= \tilde{ H}^{(N)}_{j\ell}(\mu) =\tilde{ H}^{(N)}_{-1, j \ell} {1\over \mu} + \tilde{ H}^{(N)}_{0, j \ell} + \mathcal{O}(\mu)
\end{equation}
with the coefficient matrices $\tilde{ H}^{(N)}_{-1,j\ell}$ and $\tilde{ H}^{(N)}_{0,j\ell}$  evaluated in terms of $\tilde{\Matrix{H}}^{(N-1)}$ as 
\begin{eqnarray}
\tilde{ H}^{(N)}_{-1,j\ell} &=& -{A}^{(N-1)}_{j v_{N-1}} {A}^{(N-1)}_{\ell v_{N-1}} \frac{\tilde{ H}^{(N-1)}_{0,j v_{N-1}} \tilde{ H}^{(N-1)}_{0,\ell v_{N-1}}}{ \tilde{ H}^{(N-1)}_{1,v_{N-1} v_{N-1}}},
\nonumber\\
\tilde{ H}^{(N)}_{0,j\ell} &=&\tilde{ H}^{(N-1)}_{0,j\ell} - {A}^{(N-1)}_{j v_{N-1}} {A}^{(N-1)}_{\ell v_{N-1}} \times \nonumber\\
&& \left\{\frac{\tilde{ H}^{(N-1)}_{1,j v_{N-1}} \tilde{ H}^{(N-1)}_{0,\ell v_{N-1}}+\tilde{ H}^{(N-1)}_{0,j v_{N-1}} \tilde{ H}^{(N-1)}_{1,\ell v_{N-1}}}{\tilde{ H}^{(N-1)}_{1,v_{N-1} v_{N-1}}}\right. \nonumber\\
&&\left.- \frac{\tilde{ H}^{(N-1)}_{0,j v_{N-1}} \tilde{ H}^{(N-1)}_{0,\ell v_{N-1}} \tilde{ H}^{(N-1)}_{2,v_{N-1} v_{N-1}}}{\left[\tilde{ H}^{(N-1)}_{1,v_{N-1} v_{N-1}}\right]^2}\right\}.
\label{eq:updateLast}
\end{eqnarray}
Then, from Eq.~(\ref{eq:pseudoInverse}),  the elements of the pseudo inverse of $\Matrix{V}$ are evaluated as  
\begin{equation}
{V}^+_{j\ell} = \tilde{ H}^{(N)}_{0,j\ell}.
\end{equation}

The above procedures for computing the exact pseudo inverse $\Matrix{V}^+$ of an $N\times N$ SPDS matrix $\bf V$  are summarized in the following:
\begin{enumerate}
\item Construct a $N\times N$ matrix $\Matrix{H}^{(0)} = \mu \Matrix{I} + 
\Matrix{V}$ and its extended matrix $\tilde{\Matrix{H}}$ of $2N\times 2N$ using 
Eq.~(\ref{eq:extension}). 
\item Construct a graph $G^{(0)}$ representing $\tilde{\Matrix{H}}$ and make its adjacency matrix $\Matrix{A}^{(0)}$.
\item $n\gets 0$. 
\item Remove the last row and column of $\tilde{\Matrix{H}}^{(n)}$ and update 
the elements  related to the neighbor nodes of the node $v_n$ using 
Eq.~(\ref{eq:update}) if $n<N-1$ or Eq.~(\ref{eq:updateLast}) for $n=N-1$. This 
yields $\tilde{\Matrix{H}}^{(n+1)}$. 
\item Assign a link between every disconnected pair of the neighbor nodes of $v_n$ and eliminate the node $v_n$ and its links in $G^{(n)}$. This yields $G^{(n+1)}$. Remove the last row and column of $\Matrix{A}^{(n)}$ and update the elements related to the neighbor nodes of $v_n$ by using Eq.~(\ref{eq:adjacencyevolution}). This yields $\Matrix{A}^{(n+1)}$.
\item If $n = N-1$, stop the process, else $n\gets n+1$ and go to the step 4.
\end{enumerate}
Here we emphasize that before applying this algorithm, the indices of the 
	matrix $\Matrix{H}$ should be  rearranged such that the ordered 
	list of decimated nodes in $\tilde{\Matrix{H}}$ is given by  $v_n = 2N-n$. 
The source code implementing the proposed algorithm is 
	available in \cite{Hwang2014}. 

The space and time complexities of the algorithm are as follows. 
If the number of neighbors of the decimated nodes is $\mathcal{O}(1)$, each step in the algorithm takes $\mathcal{O}(1)$ time and the whole algorithm will take $\mathcal{O}(N)$ time as the step 4 and 5 are repeated $N$ times.
This implies that the number of non-zero elements of $\Matrix{H}$ is $\Order{N}{}$ throughout the computation and 
$\mathcal{O}(N)$ space of memory is sufficient.
On the other hand, if the number of neighbors of the decimated node is of order 
$N$ and thereby $\mathcal{O}(N^2)$ elements should be updated at the step 4 and 
5, the computation time scales as $\sim N^3$.
Also, $\Matrix{H}$ becomes dense sand $\mathcal{O}(N^2)$ memory is needed.
Therefore the computational cost of our algorithm depends critically on the 	
	network topology such that the space and the time complexity are  
	$\mathcal{O}(N)$ and $\mathcal{O}(N)$ in the best case and  
	$\mathcal{O}(N^2)$ and $\mathcal{O}(N^3)$ in the weakest case, 
	respectively. 
It depends also on the order of decimating nodes 
	while we do not explore this issue systematically here.  
In the next section, we investigate the performance of our algorithm in more 
detail, focusing on time complexity.

\section{Performance of our algorithm in computing the GMFPT}
\label{sec:performance}

The major use of our algorithm lies in its ability to compute a set of arbitrary elements of 
the  exact pseudo inverse of a large SPDS matrix, which are important in the physics context at the least as described 
in Sec.~\ref{sec:target}. Furthermore, its computing time is
 much shorter than $\mathcal{O}(N^3)$ for most matrices, as we will show in this section, which enables us to apply the algorithm to large matrices constructed from big data. 

To address the performance of the proposed algorithm specifically, we investigate the computing time $\mathcal{T}$ taken to obtain {\it all} the diagonal elements of the symmetric Laplacian matrix $\bar{\Matrix{L}}$ of diverse networks including artificial and real ones. 
As shown in Eq.~(\ref{eq:GMFPT}), the set of all the diagonal components $\{\bar{L}^+_{jj}|j=1,2,\ldots, N\}$ 
indicates the GMFPT's to all nodes, $T_{j}$'s, in a network having the symmetric Laplacian matrix $\bar{\Matrix{L}}$.

\subsection{GMFPT from the simulation of random walk}
For comparison, let us consider estimating the GMFPT $T_j$  by performing the simulation of random walk on a given sparse network of $N$ nodes and $L=\Order{N}{}$ links.
The average of the MFPT for $m$ random walkers starting at arbitrary locations 
and arriving at node $j$ gives the  {\it ensemble average}  $\langle 
T_j\rangle$. The advantage of the random walk simulation is that the required 
memory is only  $\mathcal{O}(N)$, much smaller than $\mathcal{O}(N^2)$ in the 
worst case of our algorithm.
Concerning the time complexity, it takes $\mathcal{T}\simeq \langle T_j\rangle 
N m$ to obtain all the 
GMFPT's $\{\langle T_j\rangle|j=1,2,\ldots, N\}$ from the simulation.   
The deviation of $\langle T_j\rangle$ from the exact value  $T_j$  scales as 
$T_j -\langle T_j\rangle\sim \langle T_j\rangle/\sqrt{m}$~\cite{Hwang2012a} and 
thus the higher accuracy we require, the larger number of ensembles (random 
walkers) we need to run.
Note that the algorithm proposed in this work provides the exact values $T_j$'s and their 
higher moments can be also obtained exactly~\cite{Hunter2014}.
We simply require that the relative error ${T_j -\langle T_j\rangle\over \langle T_j\rangle }$ should be statistically less than ${1\over \sqrt{\langle T_j\rangle}}$, which leads to the requirement that the number of ensemble 
should be larger than $\langle T_j\rangle$, i.e., $m\geq \langle T_j\rangle$. 
The total simulation time is then given by 
\begin{equation}
\mathcal{T}\sim \max_j \{\langle T_j\rangle^2\} N.
\end{equation} 
It is known that 
\begin{align}
\max_j\{\langle T_j\rangle\} \sim \left\{
\begin{array}{ll}
N^{2/d_s} & \ {\rm  for} \  d_s<2,\\
N & \ {\rm for} \ d_s>2
\end{array}
\right.
\label{eq:Tj}
\end{align}
 with $d_s$ the spectral dimension of the underlying network ~\cite{Tejedor2009,Meyer2011,Hwang2012a}.  Therefore the whole simulation time needed to obtain such accurate  ensemble averages $\langle T_j\rangle$'s for all $j=1,2,\ldots, N$ as the relative error being less than $1/\sqrt{\langle T_j\rangle}$ scales  as 
\begin{align}
\mathcal{T} &\sim  N^z \label{eq:dynamicexponent}\\
z &= \left\{
\begin{array}{ll}
{4\over d_s}+1 & \ {\rm for} \ d_s<2, \\
3 & {\rm for} \ d_s>2.
\end{array}
\right.
\label{eq:Tsim}
\end{align}
It is remarkable that the simulation time  decreases with the spectral dimension; A very long simulation is needed for estimating $T_j$'s in networks of low dimensionality. For instance, $\mathcal{T} \sim N^5$ for $d_s=1$ and $N^3 < \mathcal{T}< N^5$ for $1<d_s<2$.  Such long simulations are not available practically for large $N$. Our algorithm gives the exact values of $\{T_j\}$ within $\mathcal{O}(N^3)$ time even in the worst case. Moreover, in contrast to the simulation time in Eq.~(\ref{eq:Tsim}), the computing time $\mathcal{T}$ of the algorithm turns out to be short for networks of low dimensionality.

\subsection{Computing time for GMFPT in model networks}

\begin{figure}
\includegraphics[width=7cm]{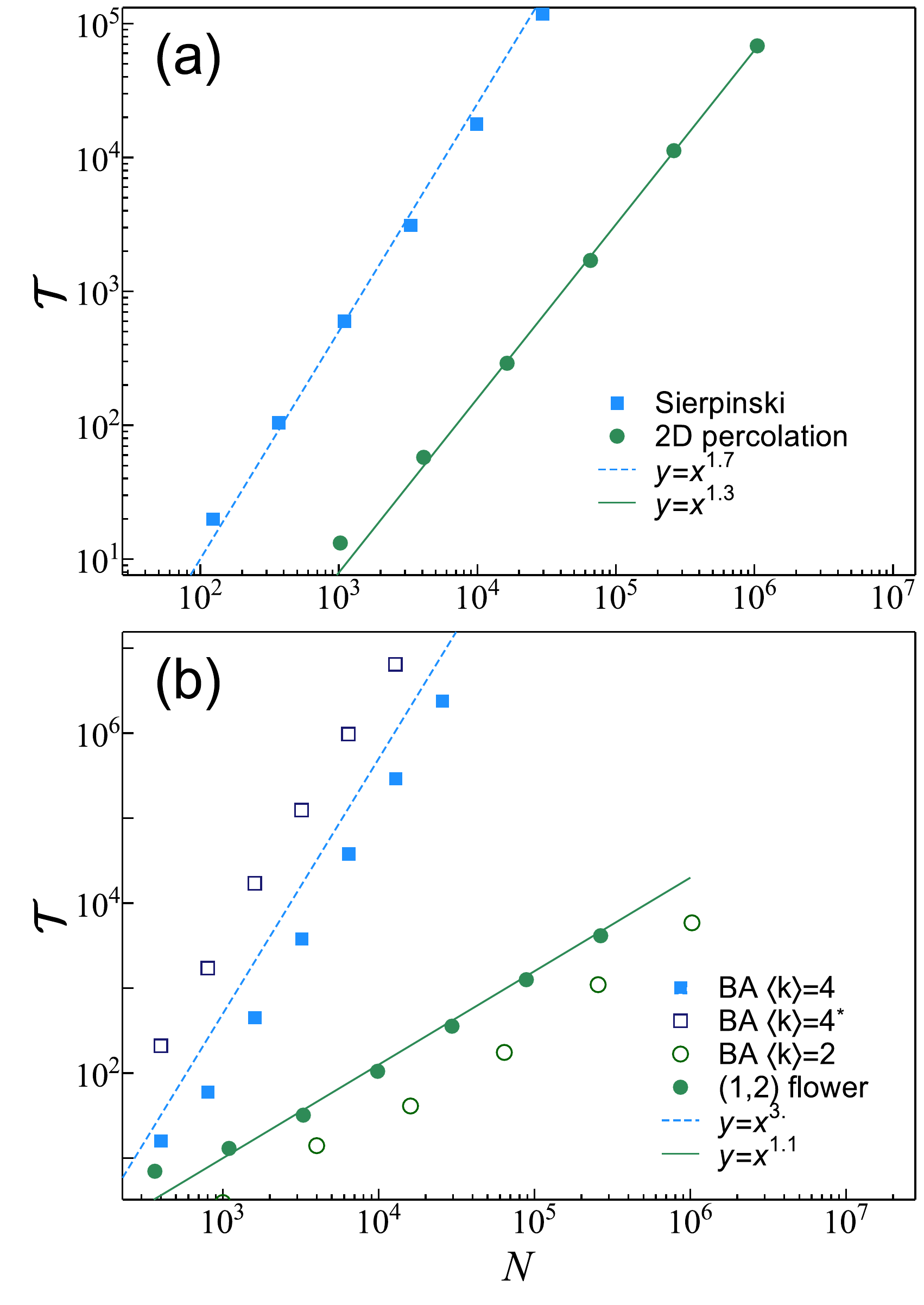}
\caption{(Color Online) Scaling of the computing time $\mathcal{T}$ for the 
generalized MFPT in model networks of $N$ nodes. 
The model networks are (a) the Sierpinski gasket and 2D critical percolation 
cluster and (b) the BA model network with the mean degree $\langle k\rangle=2$ 
and  $\langle k\rangle=4$ and (1,2)-flower networks. 
For comparison, we also draw the result for the BA model
	($\square$) with the same mean degree $\Braket{k}=4$ but using the 
	conventional eigendecomposition.
We have not shown the scaling of computing times obtained by using the 
conventional method for other cases but the BA model with 
$\braket{k}=4$. 
Because, they all behave as $\mathcal{T} \sim \mathcal{O}(N^3)$.
}
\label{fig:time}
\end{figure}

The performance of our algorithm varies with the network topology. 
In Fig.~\ref{fig:time}, we present the computing time $\mathcal{T}$ of the 
GMFPT's as a function of the number of nodes $N$ for various model networks 
such as the Sierpinski gasket($d_f = \ln 3/\ln 2$ and $d_s=2 \ln 3/\ln 
5$)~\cite{Avraham2005}, two-dimensional (2D) percolation clusters at the 
critical point($d_f = 91/48$  and $d_s=1.32$)~\cite{Rammal1984}), the 
Barab\'asi-Albert  (BA) model networks with a power-law degree 
distributions~\cite{Barabasi1999}
($d_f\to\infty, d_s=4/3$ for $\braket{k}=2$ and $ d_f\to\infty, d_s \to\infty$ for 
$\braket{k}>2$){\cite{Durhuus2007,Guclu2007,Samukhin2008}, and the 
$(1,2)$ flower 
networks ($d_f =\infty, d_s= 2 \ln 3 / \ln 2$) 
\cite{Rozenfeld2007,Rozenfeld2007a,Hinczewski2006}, where $d_f (d_s)$ is the 
fractal (spectral) dimension.
If we measure the scaling exponent $z$ introduced in Eq.~(\ref{eq:dynamicexponent}) also for the computing time 
$\mathcal{T}$ of our algorithm in each network, the exponent $z$ turns out to be different as shown in Fig.~\ref{fig:time}. 

We can classify those studied networks according to their fractal dimensions or 
	the spectral dimensions.
The Sierpinski gasket and the 2D critical percolation 
cluster have finite fractal dimensions and other networks are not fractal, 
having infinite fractal dimensions. The spectral dimension is infinite in the 
BA model networks with $\langle k\rangle=4$;
	however, it is finite between $1$ and $2$, for other networks. 

The Sierpinski gasket and the 2D critical percolation cluster 
have their node degree bounded. Given such finite node degrees, the computing 
time of step 4 and 5  at each iteration in the algorithm in 
Sec.~\ref{sec:algorithm} is expected to be  $\mathcal{O}(1)$ unless lots of 
fill-ins are generated during renormalization. The scaling exponent $z$ in 
Eq.~(\ref{eq:dynamicexponent}) is indeed $z = 1.7$ and $z=1.3$ for the 
Sierpinski gasket and the 2D percolation cluster, respectively, in 
Fig.~\ref{fig:time} (a). 
Both  are far smaller than $z=3$ of the worst case.
This suggests that it affects the time complexity of 
our algorithm whether the degree is bounded or not.  
The Sierpinski gasket is constructed recursively and shows the self-similarity of fractal structures. The minimum-degree node is the oldest one in the Sierpinski gasket, which generates fill-ins. The structure of the 2D percolation cluster is not deterministic but random due to the removal of randomly-selected sites during the course of its construction from a regular 2D lattice. The shorter computing time $\mathcal{T}$ in the percolation cluster  implies that a smaller number of fill-ins are generated than in the Sierpinski gasket by the minimum-degree ordering.

In Fig.~\ref{fig:time} (b), we present the computing time of the GMFPT in two scale-free (SF) networks: the BA model networks with $\langle k\rangle=2$ and $4$ and the (1,2) flower networks. They are not fractal. The node degrees are not bounded and therefore the computing time of step 4 and 5 at each iteration can be long. The scaling exponent $z$ of the computing time 
is expected to be larger than the networks with bounded degrees. However, $\mathcal{T}$ is the shortest for the $(1,2)$ flower networks among the four classes of  networks  in Fig.~\ref{fig:time}. On the contrary, the BA model networks with $\langle k\rangle=2L/N=4$ show the longest computing time. The origin of such striking difference  can be found  in  their network structures. The flower networks are constructed in a recursive way with the youngest node having the minimum degree. Eliminating the minimum-degree nodes is thus exactly the reverse  of the original construction process and does not create any fill-in. Furthermore, every node has only two neighbors at the moment of elimination, which leads to the almost linear scaling ($z=1$) of the computing time as shown in Fig.~\ref{fig:time} (b). On the other hand, the BA model networks are random networks displaying power-law degree distributions, for which  lots of fill-in's can be created during renormalization. These BA networks with $\langle k\rangle=4$ become almost completely connected already in the early stage of evolution and thus the step 4 and 5 take $\mathcal{O}(N^2)$ time at each iteration, leading to $z=3$, the largest value of $z$ possible in our algorithm. It should be also noted that the 
BA networks with $\langle k\rangle=2$ have the computing time scale in a similar way to that of the $(1,2)$ flower 
networks, much shorter than that of the BA networks with $\langle k\rangle=4$.  
Their difference is that 
a BA network with $\langle k\rangle=2$ is of tree structure and has a finite spectral dimension ($d_s=4/3$) 
in contrast to the BA networks with $\langle k\rangle=4$ that have loops and $d_s\to\infty$.

In spite of such varying behaviors of the computing time from network to network, the performance of our algorithm 
in computing the GMFPT is better than that of the the random-walk simulation in all the studied networks. 
Interestingly, in contrast to the  simulation time, the computing time of the algorithm tends to be shorter in networks of low dimensionality than those of high dimensionality, characterized by $d_f$ and $d_s$, meaning that  the algorithm is particularly useful for the networks of low dimensionality. We also observe that the structural characteristics other than dimensionality, such as hierarchy and randomness, and the ordering scheme for eliminating nodes may affect the computing time and even the scaling exponent $z$. It has been shown that there  exists an ordering which provides the upper bound of the number of fill-ins less than $\mathcal{O}(N^{1/4} (\log{N})^{7/2})$  and therefore $\mathcal{T}\sim N^{5/4} (\ln N)^{7/2}$ for a given sparse matrix~\cite{George2011}.
Therefore the computing time can be reduced drastically if  the optimal ordering can be found and applied.  
Various ordering schemes other than the minimum-degree one can be found in e.g., Ref.~\cite{Karypis2009}.

The conventional eigendecomposition method based on the QR algorithm \cite{Press2007} can be applied to obtain 
the GMFPT if the size of the Laplacian matrix is not so large. 
The conventional method takes $\mathcal{O}(N^3)$ time, whether the matrix is sparse or not \cite{Press2007}.  Its computing time for the BA model networks with $\langle k\rangle=4$ is presented for $N\lesssim 10^4$ in Fig.~\ref{fig:time} (b).  While our algorithm shows the worst performance, $\mathcal{O}(N^3)$, for the BA networks with $\langle k\rangle=4$ in Fig.~\ref{fig:time},  it is shown to be better than the conventional method  
with the ratio of the computing times of the two algorithms ${\mathcal{T}_{\rm 
ours}\over \mathcal{T}_{\rm conv.}}\simeq  0.03 \pm 0.015 $ almost constant in 
our simulation range  	$ 800 \leq N \leq 12800$.
It is obvious that our algorithm outperforms the  conventional method for other networks, 
for which our algorithm shows $\mathcal{O}(N^z)$ time complexity 
 with $z<3$ but the conventional one shows $\mathcal{O}(N^3)$ one. 
Given that the computing time of the conventional method is $10^7$ ms ($2.7$ hours) for the BA networks 
with $\langle k\rangle=4$ and $N=10^4$,  
one can see that  it amounts to $2700$ hours $\approx 115$ days for $N=10^5$ and thus 
the conventional method does not work for the BA networks with $N= 10^5$.

We should mention that all the computations, whatever algorithms we use, and all the simulations have been performed   in the same identical computer equipped with Intel i7, 3.4Ghz CPU and 8 GB memory. 
In compiling the source code C{}\verb!++!, we  switching on the gcc's compiler options 
``-O3 -ffast-math'' for optimization.
Especially, for the computation by the conventional eigendecomposition method, we used the implementation of the Eigen library,
	 which is believed to be one of the most efficient linear algebra library \cite{Guennebaud2010}.

\begin{figure}
\includegraphics[width=8.5cm]{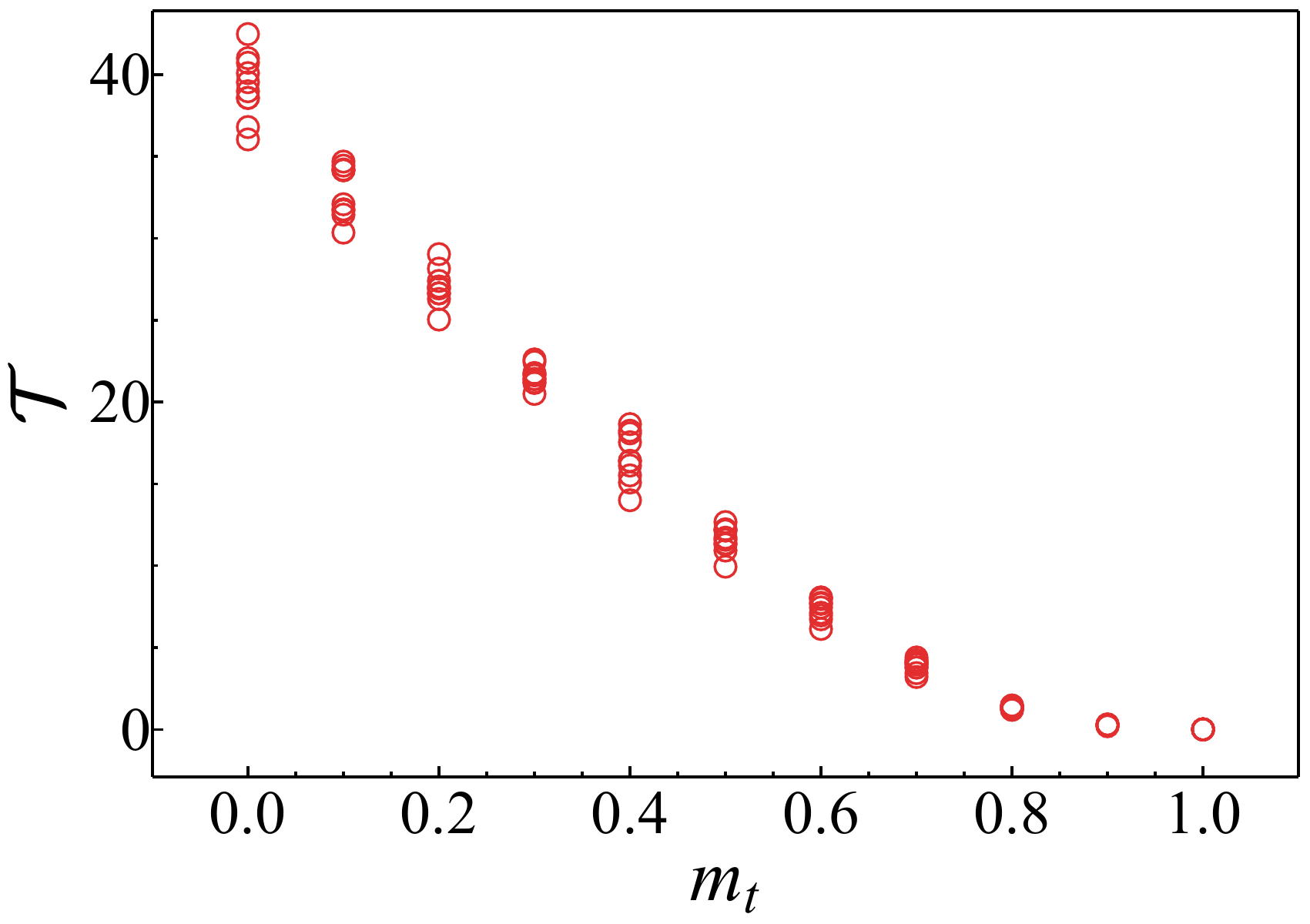}
\caption{(Color Online) 
The computing time $\mathcal{T}$ (in seconds) for the GMFPT in the modified BA networks of $N=10^4$ and $\langle k\rangle=4$ with the clustering coefficient controlled by $m_t$. The larger $m_t$ is, the larger the clustering coefficient is.  For each given value of $m_t$, 10 networks are sampled and their computing times are plotted.}
\label{fig:holme}
\end{figure}

Finally, we also investigate the dependence of network clustering on the 
computing time of our algorithm. 
The clustering coefficient of a network \cite{Watts1998} quantifies the likelihood that two neighbors of a node are also connected to each other. The number of fill-ins is therefore 
expected to be smaller  for a network with high clustering than that with low clustering if both  have the 
same number of nodes and links in the beginning. 
For a variant of the BA model  with 
a parameter $m_t$ controlling the clustering coefficient~\cite{Holme2002}, 
the computing time of the GMFPT is indeed  decreasing with increasing the clustering 
coefficient ($m_t$) in the model network of $N=10^4$ nodes and $\langle k\rangle=4$  
as shown in Fig~\ref{fig:holme}.

\subsection{Computing time for GMFPT in real networks}

The scaling behaviors of the computing time, $\mathcal{T}\sim O(N^z)$ with 
$z<3$, identified in most of the studied artificial networks, suggest that our 
algorithm can be useful in analyzing the Laplacian matrices 
of large real-world systems. 
We constructed the Laplacian matrices $\bar{\Matrix{L}}$ of one email-communication network, the subgraphs of the World-Wide-Web (WWW), and two road networks in the United States, 
all archived in the Stanford Large Network Dataset Collection~\cite{SNAP}. These selected networks commonly have a very large number of nodes, $N$ ranging between $2\times 10^5$ and $2\times 10^6$ and the mean degree $\langle k\rangle$ between $2$ and $16$.
The properties of those real-world networks and the computing time of the GMFPT's $\{T_j\}$ by our algorithm are shown in \cref{real}.
Most importantly, we found that 
our algorithm can obtain all the GMFPT's in five minutes for email network and road networks 
and in one or two hours for the WWW. Such fast computation of the pseudo inverse of matrices of size 
millions by millions strongly suggests that our algorithm  can be applied to the analysis of diverse big-data systems 
demanded increasingly in this era.  
\footnote{We also tried but failed to obtain  the GMFPT's  in the collaboration network ``com-DBLP'' of 
334863 nodes and $\braket{k}=5.53$, owing to insufficient memory for the 
increasing number of non-zero elements during renormalization. As our algorithm works for 
larger networks, we expect that the optimal ordering for this network, other 
than the minimum-degree ordering, should 
enable the computation.}
 
 Also, it is interesting that the computing time $\mathcal{T}$'s are scattered  seemingly regardless of the size $N$; the computing time is shorter  for road networks of more than one million nodes than for the WWW consisting of less than half million nodes.  
This is not explained by their clustering coefficients that would predict the longer computing time for the networks 
of low clustering as in Fig.~\ref{fig:holme}.  The trace of the pseudo inverse ${\rm Tr}\, \bar{\Matrix{L}}^+/N$ is related to the GMFPT by Eq.~(\ref{eq:GMFPT}) and is given in \cref{real}. The road networks show larger values of ${\rm Tr}\, \bar{\Matrix{L}}^+/N$ than the WWW.  From Eq.~(\ref{eq:Tj}), we can conjecture that the spectral dimensions $d_s$ of the road networks are smaller than those of the WWW and suspect that the smaller values of $d_s$ may be related to such  short computing time in the road networks.  We have already seen that the computing time is short in the model networks of low dimensionality. 

\begin{table}
\begin{tabular}{l*{6}{c}}
Network             & $N$ & $L$ & $\langle k\rangle$ & C.C. & ${\rm Tr} \bar{\Matrix{L}}^+ / N$ &$\mathcal{T}$(sec) \\
\hline
Email-EuAll 		&  224832 &  339925 & 3.02& 0.07 & 21.3529 &87.5  \\
web-Stanford        &  255265 & 1941926 & 15.2& 0.60 & 18.7769 & 2833  \\
web-NotreDame       &  325729 & 1090108 & 6.69&  0.23 & 39.5499 & 6608  \\
roadNet-CA      	& 1957027 &	2760388 & 2.82&  0.05 &  916.898 & 351  \\
roadNet-TX 	        & 1351137 &	1879201 & 2.78& 0.05 & 862.147 & 165  \\
\end{tabular}
\caption{Listed are the number of nodes ($N$), the number of links ($L$), the 
mean degree $\langle k\rangle=2L/N$, the clustering coefficient (C.C.), the 
trace of the symmetric Laplacian matrix ${\rm Tr} \, \bar{\Matrix{L}}^+/N$, and 
the computing time ($\mathcal{T}$) for the GMFPT are given for each network.  }
\label{real}
\end{table}
		
\section{Summary and discussion}
\label{sec:summary}

In this work, we proposed an algorithm that computes a set of arbitrary elements of the exact pseudo inverse of a class of singular matrices, which we call the SPDS matrices. This class of matrices play the role of the time-evolution operators in the Markov chains satisfying the detailed-balance condition and 
the elements of their pseudo inverse contain important information such as the MFPT and the correlation 
function.   Therefore fast and efficient algorithms of computing the elements of the pseudo inverse of the SPDS 
matrices can be greatly useful for analyzing the dynamics of large 
complex systems in this big-data era. Our algorithm consists of the steps of 
decimating the variables in the Gaussian integral and renormalizing the 
Hamiltonian matrix repeatedly. The algorithm runs very fast occupying little 
space of memory in many cases, which  enables us to apply the algorithm to 
large-sized singular matrices,
	e.g., of size millions by millions, capturing the dynamics of large complex 
	systems. 

The optimal order of decimating nodes, once found, would greatly reduce the computing time of our algorithm,  which needs further investigation for practical applications. We have shown that our algorithm allows us to obtain 
 the diagonal elements of the pseudo inverse of the Laplacian matrices of real-world networks such as the WWW, email-communication, and road networks of millions of nodes within minutes or a few hours, which suggests strongly 
the potential of our algorithm in analyzing the relaxation processes  in  big-data systems. 

\begin{acknowledgments}
This work was supported by the National Research Foundation of Korea (NFR)
grants funded by the Korean Government (MSIP and MEST) (No.~2010-0015066 (BK) and No.~2012R1A1A2005252  (DSL)).
\end{acknowledgments}
\appendix

\section{A faster algorithm computing $\operatorname{Tr} \Matrix{V}^{+}$}
\label{appendix2}

For computing quantities like the roughness $w$ defined in Eq.~(\ref{eq:roughness}), it is only $\operatorname{Tr}\, \Matrix{V}^{+}$ that is needed. In such a case,  the auxiliary variables $|J\rangle$ are not needed and nor is the extended matrix $\tilde{\Matrix{H}}$, which greatly reduces the running time of the algorithm.

Let us consider a SPDS matrix $\Matrix{V}$ and the coupling matrix $\Matrix{H}(\mu)= \mu \Matrix{I} + \Matrix{V}$. 
Since $\operatorname{Tr}  \Matrix{V}^{+} \equiv \sum_{\ell=2}^{N} \frac{1}{\lambda_\ell}$ with $\lambda_\ell$'s being the eigenvalues of $\Matrix{V}$, one can use the expansion of $\det\Matrix{H}$ as 
\begin{align}
\det \Matrix{H} &= \prod_{n=1}^{N} (\mu + \lambda_n) \nonumber\\
&= a_N \mu^N + a_{N-1} \mu^{N-1} + \cdots + a_2 \mu^2 + a_1 \mu.
\label{eq:a1a2}
\end{align}
with 
\begin{equation}
\operatorname{Tr} \Matrix{V}^{+} = \frac{a_2}{a_1}.
\end{equation}
Considering the application of the procedures in Sec.~\ref{sec:algorithm} to $\Matrix{H}$, not to $\tilde{\Matrix{H}}$, one finds that 
\begin{equation}
\det \Matrix{H} = \prod_{n=0}^{N-1} {H}^{(n)}_{v_n v_n}.
\end{equation} 
Using the expansion of $\Matrix{H}^{(n)}_{v_n v_n}$ in terms of $\mu$ as 
\begin{equation}
\Matrix{H}^{(n)} = \left\{
\begin{array}{ll}
\Matrix{H}^{(n)}_0 + \Matrix{H}^{(n)}_1 \mu + \Order{\mu}{2}  & \ (0\leq n<N-1)\\
\Matrix{H}^{(N-1)}_1 \mu  + \Matrix{H}^{(N-1)}_2 \mu^2 + \Order{\mu}{3}  & \ (n=N-1)
\end{array}
\right.
\end{equation}  
one can obtain $a_1$ and $a_2$ in Eq.~(\ref{eq:a1a2}).  
Finally, $\operatorname{Tr} \Matrix{V}^{+}$ is evaluated as 
\begin{equation}
\operatorname{Tr} \Matrix{V}^{+} = \sum_{n=0}^{N-2} \frac{{H}_{1,v_n v_n}^{(n)}}{{H}_{0,v_n v_n}^{(n)}}+\frac{{H}_{2,v_{N-1} v_{N-1}}^{(n)}}{{H}_{1,v_{N-1} v_{N-1}}^{(n)}}.
\end{equation}
.

\bibliography{InverseMatrixGaussianIntegration} 

\end{document}